\documentclass[11pt]{article}
\usepackage[margin=1in]{geometry}
\usepackage{amsmath,amssymb,amsthm}
\usepackage{graphicx,booktabs,xcolor}
\usepackage[numbers]{natbib}
\usepackage{algorithm,algpseudocode}
\usepackage{placeins}
\usepackage{hyperref}
\hypersetup{colorlinks=true,linkcolor=blue!50!black,citecolor=blue!50!black,urlcolor=blue!50!black}
\newcommand{\builtdate}{9 October 2026}
\newcommand{\totalruns}{115{,}696}
\newcommand{\ncampaigns}{16}

\newcommand{\CsevenHardenLowest}{142}

\newcommand{\CsevenHardenSolved}{125}
\newcommand{\CsevenGrowLowest}{144}

\newcommand{\CsevenGrowSolved}{129}
\newcommand{\CsevenRigidLowest}{155}

\newcommand{\CsevenRigidSolved}{133}

\newcommand{\CsevenSignRigid}{6 vs 14}
\newcommand{\CsevenSignPRigid}{0.12}

\newcommand{\CsevenInstances}{188}
\newcommand{\CsevenUnsolved}{43}
\newcommand{\CsevenSquInSquHardenLowest}{27}
\newcommand{\CsevenSquInSquHardenSole}{5}
\newcommand{\CsevenSquInSquRigidLowest}{23}

\newcommand{\CsevenSquInSquN}{28}

\newcommand{\CsevenTriInTriHardenLowest}{18}

\newcommand{\CsevenTriInTriRigidLowest}{26}

\newcommand{\CsevenTriInSquHardenLowest}{14}

\newcommand{\CsevenTriInSquRigidLowest}{22}

\newcommand{\CsevenHexInSquHardenLowest}{15}

\newcommand{\CsevenHexInSquRigidLowest}{16}

\newcommand{\CsevenHexInSquN}{18}

\newcommand{\CubesHardenTwelve}{2.942809}
\newcommand{\CubesRigidTwelve}{3.000000}
\newcommand{\CubesGrowTwelve}{2.947506}

\newcommand{\CtauOne}{7}
\newcommand{\CtauHalf}{15}
\newcommand{\CtauQuarter}{18}
\newcommand{\CtauZero}{17}
\newcommand{\CtauN}{40}
\newcommand{\CtauSquares}{3, 4, 2, 0}
\newcommand{\CtauTriangles}{2, 7, 8, 10}

\newcommand{\CbudgetN}{20}
\newcommand{\PortRigidSixtyFour}{138}
\newcommand{\PortRigidHarden}{139}
\newcommand{\PortRigidGrow}{137}
\newcommand{\PortSign}{4 vs 3}
\newcommand{\PortP}{1}

\newcommand{\BootDRH}{3 to 24}

\newcommand{\BootDRHPos}{99\%}

\newcommand{\BootDRGPos}{99\%}
\newcommand{\BootDGH}{-13 to 9}

\newcommand{\EligHarden}{97\%}

\newcommand{\EligRigid}{100\%}

\newcommand{\EligPc}{56\%}
\newcommand{\RangeSmallHarden}{54}
\newcommand{\RangeSmallGrow}{54}
\newcommand{\RangeSmallRigid}{56}
\newcommand{\RangeSmallSa}{56}
\newcommand{\RangeSmallPc}{47}
\newcommand{\RangeSmallN}{58}
\newcommand{\RangeMidHarden}{50}
\newcommand{\RangeMidGrow}{57}
\newcommand{\RangeMidRigid}{52}
\newcommand{\RangeMidSa}{43}
\newcommand{\RangeMidPc}{20}
\newcommand{\RangeMidN}{70}
\newcommand{\RangeLargeHarden}{38}
\newcommand{\RangeLargeGrow}{33}
\newcommand{\RangeLargeRigid}{47}
\newcommand{\RangeLargeSa}{26}
\newcommand{\RangeLargePc}{5}
\newcommand{\RangeLargeN}{60}

\newcommand{\RefEvalsTwenty}{438{,}032}
\newcommand{\RefRatioTwenty}{1.05}
\newcommand{\RefEvalsThirty}{696{,}566}
\newcommand{\RefRatioThirty}{1.07}

\newcommand{\SnapHLower}{17}
\newcommand{\SnapXLower}{36}

\newcommand{\SnapP}{0.013}

\newcommand{\SnapRXLower}{20}
\newcommand{\SnapRRLower}{20}

\newcommand{\SnapSquSqu}{6--0}

\newcommand{\SnapTriTri}{0--8}

\newcommand{\SnapTriSqu}{6--14}

\newcommand{\AreaHLower}{20}
\newcommand{\AreaXLower}{31}

\newcommand{\AreaP}{0.16}

\newcommand{\AreaRXLower}{15}
\newcommand{\AreaRRLower}{14}

\newcommand{\AreaSquSqu}{5--1}

\newcommand{\AreaTriTri}{1--8}

\newcommand{\BatN}{28}
\newcommand{\BatHReached}{49}

\newcommand{\BatHSome}{4}
\newcommand{\BatRReached}{46}

\newcommand{\BatRSome}{2}
\newcommand{\BatSqElevenH}{2}
\newcommand{\BatSqElevenR}{0}
\newcommand{\BatSqTwentyNineH}{2}
\newcommand{\BatSqTwentyNineR}{0}
\newcommand{\BatTriTwentyNineH}{1}
\newcommand{\BatTriTwentyNineR}{3}
\newcommand{\LongN}{14}
\newcommand{\LongHardenLowest}{13}

\newcommand{\LongHardenReached}{12}
\newcommand{\LongGrowLowest}{7}

\newcommand{\LongRigidLowest}{8}

\newcommand{\LongRigidReached}{8}
\newcommand{\LongSaLowest}{9}

\newcommand{\LongSaReached}{9}
\newcommand{\LongPcLowest}{6}

\newcommand{\LongHardenOnlyReachedNewNs}{$n=17$ and $n=26$}
\newcommand{\LongNewHLower}{3}
\newcommand{\LongNewRLower}{0}
\newcommand{\LongNewP}{0.25}
\newcommand{\LongNewN}{11}
\newcommand{\LongHLower}{6}
\newcommand{\LongRLower}{0}
\newcommand{\LongP}{0.031}

\newcommand{\LongHardenOnlyReachedNs}{17, 19, 26, 29}
\newcommand{\LongSaOnlyReachedNs}{27}

\newcommand{\BudOneH}{7}
\newcommand{\BudOneR}{2}
\newcommand{\BudOneP}{0.18}

\newcommand{\BudThreeH}{4}
\newcommand{\BudThreeR}{3}

\newcommand{\BudMoreH}{6}
\newcommand{\BudMoreR}{2}

\newcommand{\BudLvMLH}{3}
\newcommand{\BudLvMMH}{3}

\newcommand{\BudGainH}{4}
\newcommand{\BudLossH}{3}
\newcommand{\BudLvMLR}{3}
\newcommand{\BudLvMMR}{2}

\newcommand{\BudGainR}{5}
\newcommand{\BudLossR}{0}
\newcommand{\BudN}{28}

\IfFileExists{decision.tex}{\newcommand{\DecN}{60}
\newcommand{\DecPilotLower}{2}
\newcommand{\DecRigidLower}{3}

\newcommand{\DecP}{1}
\newcommand{\DecHLower}{3}
\newcommand{\DecHRLower}{17}
\newcommand{\DecHP}{0.0026}
\newcommand{\DecPicked}{20}
\newcommand{\DecReachedRigid}{40}
\newcommand{\DecReachedHarden}{32}
\newcommand{\DecReachedPilot}{41}
\newcommand{\DecTPicked}{3}
\newcommand{\DecTChanged}{17}
\newcommand{\DecTLower}{2}
\newcommand{\DecTRLower}{3}
}{}
\IfFileExists{refine.tex}{\newcommand{\RefN}{28}
\newcommand{\RefRawHarden}{6}
\newcommand{\RefRawGrow}{19}
\newcommand{\RefRawRigid}{3}
\newcommand{\RefRawSa}{1}
\newcommand{\RefRawPc}{0}
\newcommand{\RefEqHarden}{23}
\newcommand{\RefEqGrow}{20}
\newcommand{\RefEqRigid}{19}
\newcommand{\RefEqSa}{17}
\newcommand{\RefEqPc}{4}
\newcommand{\RefPolHarden}{24}
\newcommand{\RefPolGrow}{20}
\newcommand{\RefPolRigid}{18}
\newcommand{\RefPolSa}{14}
\newcommand{\RefPolPc}{2}
}{}

\newcommand{\MechSquSquHardenOrder}{0.53}

\newcommand{\MechSquSquRigidOrder}{0.81}

\newcommand{\MechTriTriHardenMove}{0.53}

\newcommand{\MechTriTriRigidMove}{0.28}

\newcommand{\MechSoleN}{5}
\newcommand{\MechSoleTiltH}{0.45}
\newcommand{\MechSoleTiltR}{0.28}
\newcommand{\MechSoleMore}{5}
\newcommand{\MechTriFailN}{4}
\newcommand{\MechTriFailMoveH}{0.54}
\newcommand{\MechTriFailMoveR}{0.30}

\title{Soft-to-Rigid: Shape Continuation as an Alternative Search\\ for Packing Congruent Shapes}
\author{Yohei Nakajima\thanks{Experiments, code and verification were developed with the assistance of Claude (Anthropic); see the acknowledgements.}}
\date{\builtdate}

\begin{document}
\maketitle

\begin{abstract}
Searches for dense packings of congruent shapes move rigid pieces. Hardening starts each piece as its inscribed disk and deforms it into the target shape while the container is compressed. We ask whether it is a search worth running alongside existing ones, and test it at matched budget and comparable tuning against rigid starts, growth, simulated annealing and perturbation--compression on seven families of squares, triangles, hexagons and disks in squares, triangles and circles (\CsevenInstances{} instances), scoring each method by the lowest container size it reaches. Hardening is not a better default: tuned rigid starts reach the lowest size most often. It is a different search. On squares in a square it reaches packings the other four methods miss, including Trump's $n=11$ and Wainwright's $n=19$ in 32 runs per method, and in a targeted follow-up campaign at ten times the budget (16 runs per method) it is the only one to reach the best known packings for \LongHardenOnlyReachedNewNs, while annealing alone reaches $n=\LongSaOnlyReachedNs$. In exploratory ablations, compressing disks and then switching to polygons at once, or growing rigid polygons along hardening's area schedule, shows no detectable difference from rigid starts, and hardening's unique packings contain more tilted pieces. The bias costs elsewhere: hardening is worse on triangles and on four families held out from all development, where the short-pilot policy we tested showed no detected advantage over always starting rigid. Shape continuation is an alternative search that reaches different packings, worth keeping beside rigid ones; we did not show that splitting a fixed budget between them pays, and we have no test that predicts where it helps. Search campaigns were planned before they ran, with amendments, exploratory analyses and post-hoc checks identified separately; every run is replayable and every number is generated from an event log.

\end{abstract}

\section{Introduction}
How small a container holds $n$ copies of a given shape? For unit squares in a square the question goes back to Erd\H{o}s and Graham \cite{erdosgraham}; catalogues now cover squares, triangles, hexagons and cubes in squares, triangles, circles and cubes \cite{friedmancenter,squaresproject}, and most entries are best known packings found by search rather than proved optimal. Most searches move rigid pieces: simulated annealing \cite{kirkpatrick}, perturbation and compression \cite{Gensane2004,gensane}, growing the pieces in size \cite{lubachevsky}, overlap minimisation in a shrinking container \cite{Umetani2009,Gardeyn2025}, or global solvers on exact non-overlap models \cite{Berthold2026}.

This paper changes the pieces instead. \emph{Hardening} starts every piece as the disk inscribed in it, compresses the disks under inward pressure on the container, and deforms them into the target shape while the pressure continues. It is graduated non-convexity \cite{blake} applied to geometry: a disk has no orientation, so the pieces are already densely packed when their corners appear and rotation becomes a choice. The shape family it moves along, a polygon dilated by a disk at fixed inradius, is not new: it is the spheropolygon of discrete-element modelling \cite{AlonsoMarroquin2008}, and Jin, Lu and Li moved dense periodic packings along its 3D analogue in both directions between tetrahedra and spheres \cite{Jin2015}. What we have not found is its use as an optimisation path toward a prescribed polygon inside a finite container, evaluated against tuned rigid searches for best known packings (Section~\ref{sec:related}). On squares it recovers several classical packings from random starts, and in 3D it found a packing of 12 unit cubes in a cube smaller than the best known one \cite{nakajima12}.

Here we ask whether hardening is a search worth running alongside existing ones. We compare three paths through shape space (hardening, growing, and rigid pieces on the same schedule) and two classical baselines on seven families of shapes and containers (\totalruns{} runs in \ncampaigns{} search campaigns, plus one tightening-only campaign), with matched budgets and comparable tuning effort. Since a packing search is judged by the best packing it finds, the comparison is by the \emph{lowest} container size each method reaches in a fixed number of runs, after constrained numerical tightening; the procedure being compared is the batch of runs, not a single run \cite{Birattari2005}. Our findings:
\begin{itemize}\itemsep1pt
\item Hardening is not a better default. Untuned, it looked better than rigid starts; tuned with comparable effort, rigid starts reach the lowest size most often over the seven families (Table~\ref{tab:main}).
\item It is a different search. On squares in a square it reaches the lowest size on \CsevenSquInSquHardenLowest{} of \CsevenSquInSquN{} instances and is the only method that low on \CsevenSquInSquHardenSole, including Trump's $n=11$ and Wainwright's $n=19$ packings, which rigid starts reach in none of our campaigns. In a targeted follow-up campaign at ten times the budget it is the only one of five methods to reach the best known packings for \LongHardenOnlyReachedNewNs; it also does so at $n=19$ and 29, but those hardening and rigid runs repeat an earlier campaign (Section~\ref{sec:learned}).
\item Exploratory ablations show that neither the disk-compressed start nor matched area growth reproduces hardening's pattern (Table~\ref{tab:ablate}), consistent with an effect of the gradual path; the pre-registered tests on squares in a square are inconclusive, and hardening's unique packings end with more tilted pieces (Figure~\ref{fig:mechanism}) without that isolating orientation as the cause.
\item The bias has a cost. Hardening is worse on triangles and on four families held out from all development, and the pilot policy we tested, which chooses the path per instance from 8 runs of each, is no better than always starting rigid (Table~\ref{tab:decision}). An earlier indication that it gains more from longer runs did not replicate.
\end{itemize}
Search campaigns were planned in writing before they ran; amendments, exploratory analyses and post-hoc checks are identified separately. Every run can be replayed, and every number in this paper is generated from an event log. Section~\ref{sec:matters} gives the practical rule that follows and describes downstream reuse of the earlier release.

\section{Related work}\label{sec:related}
\paragraph{Changing shape during a packing simulation.} Hard-particle physics has long changed particle shape during a run. The event-driven molecular dynamics of Donev, Torquato and Stillinger \cite{Donev2005a}, which generalises Lubachevsky--Stillinger growth \cite{lubachevsky}, explicitly allows time-dependent particle shapes. ``Digital alchemy'' \cite{vanAnders2015,Geng2019} treats shape parameters as thermodynamic variables sampled by Monte Carlo shape moves, to design shapes for a target structure. Closest to our parametrisation, Jin, Lu and Li \cite{Jin2015} take spherotetrahedra (a tetrahedron dilated by a ball, with a roundness ratio defined relative to a tetrahedron of the same inscribed sphere) and evolve dense periodic packings along that family in both directions: from tetrahedron packings toward spheres, and, in an ``inverse evolution'', from sphere packings toward tetrahedra. Superdisk and superball families \cite{Jiao2008,Jiao2009}, truncation families \cite{Engel2012} and rounded cubes in spherical confinement \cite{Wang2018} interpolate between round and faceted shapes, but packings are computed separately at each shape. Minkowski rounding is standard in discrete-element modelling as spheropolygons \cite{AlonsoMarroquin2008,GalindoTorres2010}, where it smooths contact geometry but is held fixed. Starting round and becoming faceted is therefore not new. What hardening adds is the setting: the same inradius-preserving family used as an optimisation homotopy toward a prescribed polygon inside a finite container, scored by the smallest container reached and compared against tuned rigid searches with ablations of the path. It is an instance of graduated non-convexity and continuation \cite{blake,AllgowerGeorg1990}, for which there is theory on when following a path beats a cold start and when it need not \cite{Mobahi2015,Hazan2016}.

\paragraph{Disks as a stand-in for polygons.} In operations research, inscribed circles appear as relaxations or surrogates: QP-Nest \cite{Jones2014} bounds nesting problems by enforcing non-overlap only on inscribed circles, and sparrow \cite{Gardeyn2025} builds its overlap proxy from inscribed ``poles''. In facility layout and floorplanning, modules are first modelled as circles and then converted to exact shapes in a second stage \cite{AnjosVannelli2002,Luo2008}. These use circles as a fixed surrogate or as a one-shot initial solution. We deform each piece continuously from its inscribed disk while the container is compressed, and test separately whether the disk-compressed start or the gradual path is responsible for any effect (campaign C11).

\paragraph{Packing in containers.} Phi-functions \cite{Bennell2010,Romanova2018} and separating-line programs \cite{Peralta2018} give exact non-overlap models; our tightening step is a separating-line program of this kind. Overlap minimisation in a shrinking container dominates irregular strip packing \cite{Umetani2009,Imamichi2009,Egeblad2007,Gardeyn2025}, with jostling as a constructive alternative \cite{Dowsland1998}; analytical placers replace overlap with a smooth density penalty of increasing weight \cite{Lu2015}; and gradient descent on poses with an adaptively shrinking container has recently been applied to irregular 3D objects \cite{Gupta2026}. Our energy is of this family; what differs is that the pieces' shapes change along the way.

\paragraph{Records.} Best known packings of equal pieces come from Friedman's survey and Packing Center \cite{friedmansurvey,friedmancenter}, the Squares Project \cite{squaresproject} and Packomania \cite{packomania}. Record-finding methods include billiards and growth \cite{lubachevsky,LubachevskyGraham1997}, perturbation--compression \cite{Gensane2004,gensane}, perturbation-based thresholding search \cite{Lai2023}, periodic adaptive-shrinking-cell and divide-and-concur optimisation \cite{TorquatoJiao2009,Kallus2010}, global MINLP solvers \cite{Berthold2026}, and LLM-driven program search \cite{Novikov2025,Georgiev2025,hyra}. Berthold et al.\ \cite{Berthold2026} solve regular polygons in regular polygons and Platonic solids in Platonic solids for $n\le30$; their supplement lists 12 cubes at $2.94281$ (Friedman's 1998 value), and Hyra \cite{hyra} later reports $2.93277$. Records for squares in a square at larger $n$ ($n\ge88$) continue to fall to an iterative ``shrink-hopping'' search with exact certificates \cite{squish}.

\paragraph{Evaluating stochastic searches.} A record search is used for the best packing it finds, so we compare the lowest size each method reaches in a fixed number of runs. Birattari and Dorigo \cite{Birattari2005} point out that the best of $N$ runs of an algorithm is one sample from a different algorithm, $N$ restarts of the first, and that this algorithm should be judged by its behaviour over repeated batches, not by one batch's best. We therefore treat ``32 runs of a method'' as the procedure under comparison, report repeated independent batches where we have them, and use resampling only as a sensitivity check, in the spirit of performance profiles and time-to-target plots \cite{Dolan2002,Aiex2007,Hansen2021}. Lai et al.\ \cite{Lai2023} report best, mean and success rate over 20 runs against Packomania; we add matched budgets, matched tuning effort and pre-registered decision rules.

\section{Method}\label{sec:method}
\paragraph{Pieces and paths.} A piece is a regular polygon $P$ with unit side and inradius $\rho$ (a unit square, triangle or hexagon), or a unit-diameter disk. Its \emph{shape state} is
\[
P(\gamma,\tau)=\gamma\big[(1-\tau)\,P\;\oplus\;\tau\rho\,D\big],
\]
the polygon scaled about its incenter, dilated by a disk of radius $\tau\rho$ (a spheropolygon \cite{AlonsoMarroquin2008}), and scaled by $\gamma$. At $\tau=1$ it is the inscribed disk, at $\tau=0$ the polygon, and for $\gamma\le1$ every state lies inside the polygon, so moving toward $(\gamma,\tau)=(1,0)$ only grows the piece. Two facts describe what the family does. Its support function is
\[
h_{P(\gamma,\tau)}(u)=\gamma\big[(1-\tau)\,h_P(u)+\tau\rho\big],
\]
for a unit direction $u$, so the dependence of a piece's extent on its orientation is scaled down by $(1-\tau)$: rotating a nearly round piece barely changes the energy, and orientation becomes a decision only late in the path. And since $A(P)=\tfrac12\,\mathrm{perimeter}\cdot\rho$, its area at $\gamma=1$ is
\[
A(\tau)=A(P)-\big[A(P)-\pi\rho^2\big]\tau^2 ,
\]
so a hardening piece gains the area outside its inscribed disk, $21\%$ of a square, $9\%$ of a hexagon and $40\%$ of a triangle. A \emph{path} is a schedule for $(\gamma,\tau)$ during the middle phase of a run (Algorithm~\ref{alg:run}). We compare three:
\begin{itemize}\itemsep0pt
\item \textbf{harden}: $\gamma=1$, $\tau:1\to0$ (disks harden into polygons);
\item \textbf{grow}: $\tau=0$, $\gamma:\gamma_0\to1$ (rigid polygons grow, as in Lubachevsky--Stillinger \cite{lubachevsky});
\item \textbf{rigid}: $(\gamma,\tau)=(1,0)$ throughout, on the same schedule;
\end{itemize}
and two classical baselines on rigid pieces: \textbf{sa}, Metropolis simulated annealing over poses and container size \cite{kirkpatrick}, and \textbf{pc}, perturbation--compression \cite{Gensane2004,gensane} (shrink, relax, perturb on failure). Two further paths serve as ablations: \textbf{snap} compresses disks exactly as harden does and then switches to the polygons at once ($\tau:1\to0$ in one step), and \textbf{grow-area} keeps rigid polygons whose area follows hardening's area schedule ($\tau=0$, $\gamma^2=1-(1-\pi\rho^2/A)\,t^2$ with $t:1\to0$).

\paragraph{Energy.} The container (square, equilateral triangle or circle) is centred at the origin with size $L$. For poses $(x_i,y_i,\theta_i)$,
\[
E=\sum_{i<j}\big(2r-\operatorname{sd}_{ij}\big)_+^2+\sum_{i}\sum_{v}\big(\text{wall violation of }v\big)_+^2+\mu L,
\]
where $r=\gamma\tau\rho$, $\operatorname{sd}_{ij}$ is the signed distance between the cores (exact vertex--edge distance when apart, minus the separating-axis penetration depth when overlapping) and $v$ ranges over core vertices. The term $\mu L$ is inward pressure. Positions, angles and $L$ follow momentum gradient descent with analytic gradients and decaying Gaussian noise. In 3D the same holds for cubes with quaternion orientations and the 15-axis separating test \cite{nakajima12}.

\begin{algorithm}[t]\caption{One run of a gradient path (budget factor $B$; base $n_C,n_M,n_S=2500,7500,1500$ steps).}\label{alg:run}
\small
\begin{algorithmic}[1]
\State random poses in a container of area $3.2\times$ the pieces' area; shape $\leftarrow$ path start ($\tau_0$, $\gamma_0$)
\For{$B n_C$ steps} \Comment{compress}
\State momentum step on $E$ with pressure $3\mu$, decaying noise
\EndFor
\For{$s=0,\dots,Bn_M-1$} \Comment{follow the path}
\State $(\gamma,\tau)\leftarrow$ path$(s/Bn_M)$; momentum step on $E$ with pressure $\mu$, decaying noise
\EndFor
\For{$s=0,\dots,Bn_S-1$} \Comment{settle with the final pieces}
\State momentum step on $E$ with pressure $\mu\cdot10^{-4s/Bn_S}$, no noise
\EndFor
\State legalise: scale centres apart until no pair overlaps; measure the smallest container
\end{algorithmic}
\end{algorithm}

\paragraph{Step details.} Each momentum step updates velocities as $v\leftarrow0.85\,v-0.04\,\nabla E$ (container size: $v_L\leftarrow0.85\,v_L-0.02\,(\mu+\partial E/\partial L)$), caps a piece's translation step at $0.03\sqrt2\,R$ ($R$ the circumradius) and its rotation at $0.03$ rad, then adds Gaussian noise. With noise multiplier $\nu$ (Appendix, frozen settings), the positional noise is $0.015\,\nu\,(1-u)$ during compression and $0.004\,\nu\,(1-u)$ during the path, where $u\in[0,1]$ is progress through the phase, and the angular noise is $1.5$ times the positional noise. The implementation is \texttt{engine/engine2d.js}.

\begin{figure}[t]\centering
\includegraphics[width=\linewidth]{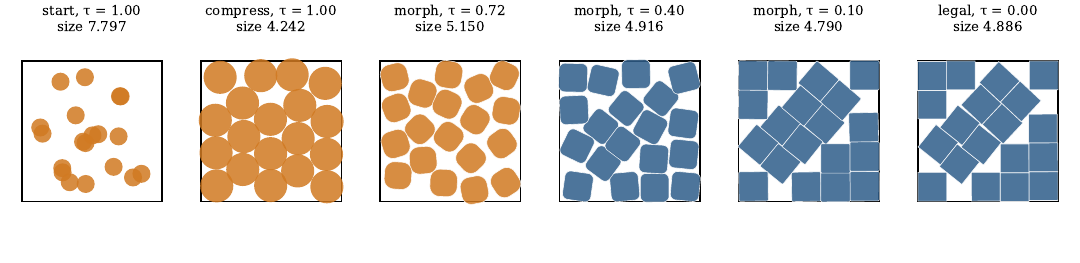}\\[2pt]
\includegraphics[width=\linewidth]{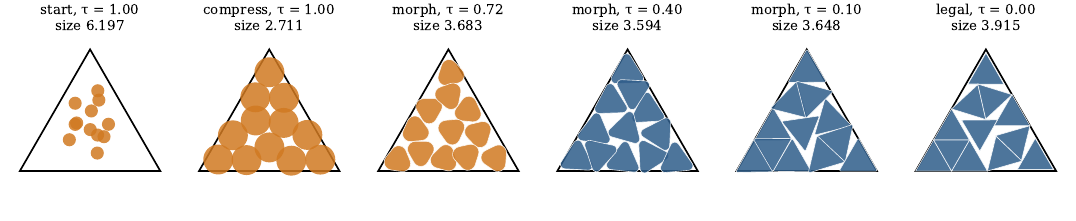}
\caption{Two hardening runs (C07). Top: 19 squares; disks are compressed, harden into squares, and settle into Wainwright's packing (side $3+4\sqrt2/3$), which no other method reached in 32 runs. Bottom: the best of 32 hardening runs for 12 triangles in a triangle ends at $3.915$, while rigid and snap starts reach the best known $3.879$; the triangles that emerge from the near-hexagonal disk arrangement keep orientations that do not tile. Every replay is on the project site.}\label{fig:strip}\label{fig:tristrip}
\end{figure}

\paragraph{From a run to a certified packing.} Penalties tolerate small overlaps, so every run is \emph{legalised}: centres are scaled apart about their mean until no pair overlaps by more than $10^{-12}$ under the separating-axis test, and the smallest container (translation free) is measured. Near-best runs are then \emph{tightened} by constrained numerical optimisation: minimise $L$ over poses and one separating line (plane) per nearby pair \cite{Peralta2018}, with every vertex inside the container, by SLSQP \cite{kraft} in trust-region outer iterations, keeping a step only if the independently legalised size decreases. Tightening is a local method and gives no optimality guarantee; it only removes the slack a penalty method leaves. A tightened packing below a catalogue value is spread to a clearance of $10^{-6}$ and checked twice: in floating point, and in exact rational arithmetic. The exact check uses rational rotations from $\tan(\theta/2)$, replaces each piece by a rational polygon scaled by $1+10^{-8}$ and proves, in $\mathbb{Q}(\sqrt3)$, that this polygon contains the true unit polygon with margin $10^{-9}$, and exhibits a rational separating direction for every pair.

\paragraph{Budgets.} A run's budget is its number of steps: every gradient path takes $11{,}500\,B$ steps. Pair evaluations then differ by path, since soft pieces have more near neighbours: at $n=20$ squares in a square a hardening run makes about $1.2\times$ the pair evaluations of a rigid run (Table~\ref{tab:diag} in the appendix), which, if anything, favours hardening. The two baselines are stopped once their pair evaluations exceed those of a reference hardening run at the same $n$ (one fixed seed, with the baseline's own pressure and noise settings); for squares in a square this reference is \RefEvalsTwenty{} pair evaluations at $n=20$ and \RefEvalsThirty{} at $n=30$, $\RefRatioTwenty$ and $\RefRatioThirty$ times the mean of the 32 tuned hardening runs. Simulated annealing then ends with a short gradient relaxation and uses about $1.1\times$ the reference; perturbation--compression stops only at the end of a relaxation and uses up to $1.8\times$. In simulator wall-clock time on one core of the study machine (Node.js; tightening excluded), a hardening run was the fastest of the five on squares in a square at $n=20$ and 30 and on every polygon family in C07 (on disks, annealing and perturbation--compression are faster) (Table~\ref{tab:diag}), so the matching does not favour it in time either.

\section{From squares to cubes}\label{sec:story}
The method was developed on unit squares in a square, where records are dense and audited \cite{squaresproject,friedmansurvey}. With the original schedule, hardening reaches Trump's packing for $n=11$ \cite{trump} after tightening (to within $10^{-14}$) in 4 of 80 random starts at three times the base budget and 2 of 80 at the base budget; rigid starts with the same schedule reach it in none of 80 at either budget (campaign C00).

The same method in 3D, with quaternion orientations and the 15-axis separating test, found a packing of 12 unit cubes in a cube of side $2.9315185$, below the previous best known $2.9327718$ \cite{hyra,nakajima12}, which had itself improved Friedman's 1998 value $2.94281$ \cite{friedmancubes} (the value the global solvers of Berthold et al.\ \cite{Berthold2026} also reach). Campaign C05 repeats the cube experiment with three gradient paths and two settings at the record's budget (Table~\ref{tab:cubes}). Hardening has the lowest side of the three paths at $n=11$, 12 and 13 under both settings, all paths reach Friedman's $n=9$ packing \cite{friedmancubes}, and with tuned settings hardening and rigid starts also reach his $n=10$ packing $2+1/\sqrt2$, which the original schedule never did. Rigid starts end at the trivial side 3 for $n\ge12$. In these 32 seeds nothing went below a catalogue value: the best side at $n=12$ is \CubesHardenTwelve{} (grow \CubesGrowTwelve, rigid \CubesRigidTwelve), while the record came from one of 192 runs. At $n=13$ hardening reaches $2.956145$, consistent with Friedman's $2.956+$ \cite{friedmancubes} to the precision printed.

\begin{table}[t]\centering\small
\begin{tabular}{rrrrrrrr}
\toprule
 & & \multicolumn{3}{c}{original settings} & \multicolumn{3}{c}{2D-tuned settings}\\
$n$ & best known & harden & grow & rigid & harden & grow & rigid\\
\midrule
9 & 2.70711 & \textbf{2.70711} & \textbf{2.70711} & \textbf{2.70711} & \textbf{2.70711} & \textbf{2.70711} & \textbf{2.70711}\\
10 & 2.70711 & 2.76777 & 2.76777 & 2.89443 & \textbf{2.70711} & 2.76777 & \textbf{2.70711}\\
11 & 2.88295 & \textbf{2.89443} & 2.91739 & 2.91899 & 2.91250 & 2.91328 & 2.94281\\
12 & 2.93277 & 2.94301 & 2.98587 & 3.00000 & \textbf{2.94281} & 2.94751 & 3.00000\\
13 & 2.95600 & 2.99497 & 3.00000 & 3.00000 & \textbf{2.95615} & 3.00000 & 3.00000\\
14 & 2.98995 & \textbf{3.00000} & \textbf{3.00000} & \textbf{3.00000} & \textbf{3.00000} & \textbf{3.00000} & \textbf{3.00000}\\
\bottomrule
\end{tabular}
\caption{Cubes in a cube (C05): lowest side over 32 runs at the record budget, after tightening. The best known column excludes our earlier $n=12$ record ($2.93152$), which these runs try to reproduce. Bold: lowest in the row.}
\label{tab:cubes}
\end{table}

Two results made the broader question worth asking. Rigid starts in these first experiments used the hardening schedule untouched, so they were never given a fair chance. And squares in a square may be special: many of its best packings tilt a few squares against their neighbours, exactly the configurations a late commitment to orientation should help with.

\section{A wider test}\label{sec:breadth}
\paragraph{Design.} Seven families: unit squares in a square, a circle and a triangle; unit triangles in a triangle and a square; unit hexagons in a square; and, as a control, unit-diameter disks in a square, where hardening changes nothing because a disk is already a disk. Best known values come from the Squares Project \cite{squaresproject}, Friedman's catalogue \cite{friedmancenter} and Packomania \cite{packomania}; truncated catalogue values (``$2.956+$'') count as reached within their printed precision. All $n$ from 2 to 30 (to 20 for hexagons) are used except five reserved for tuning, \CsevenInstances{} instances in all.

Each method was tuned twice, first by median gap on held-out instances (C01--C03; a $3\times3$ grid, extended past its edge by a fixed rule, so harden, sa and pc saw 27 settings and grow and rigid 18), then by the lowest size reached (C06; one $3\times3$ grid per method, 16 runs per setting on each of eight held-out instances of squares in a square and in a circle and triangles in a square and in a triangle, scored by the mean over instances of the lowest of the 16 runs). The second round used larger instances ($n>30$) than the test set, so its settings may be less suited to small $n$; Table~\ref{tab:range} breaks the results down by $n$. Each method then ran 32 times per instance at the matched budget of Section~\ref{sec:method}, and in each instance and method the best run (if within 2\% of the best known value) and any run within $2\cdot10^{-3}$ of it, at most three, were tightened. Every campaign's plan, including its decision rule, was committed before its first run; changes after a campaign started are recorded as amendments.

\paragraph{Result.} Table~\ref{tab:main} counts, per family, the instances on which each method's lowest size is the lowest of all five. Rigid starts reach the lowest size most often overall (\CsevenRigidLowest{} instances, against \CsevenGrowLowest{} for grow and \CsevenHardenLowest{} for harden) and reach the best known value most often (\CsevenRigidSolved{} against \CsevenGrowSolved{} and \CsevenHardenSolved{}). As a check of sensitivity to the particular seeds, we resampled each instance's 32 runs with replacement (1000 times) and recounted. Within each resample, rigid's count exceeds hardening's in \BootDRHPos{} of resamples (difference \BootDRH, 95\% range) and grow's in \BootDRGPos{}, while grow against harden goes either way (\BootDGH). This measures sensitivity to the seed pool on these instances, not uncertainty about future campaigns: a resample cannot contain a run better than the ones observed, it omits a cell's single best run with probability $(31/32)^{32}\approx0.36$, which is why every method's resampled count is lower than its observed one, and tightened sizes are fixed by the original pool rather than re-selected in each resample. Over instances solved by exactly one of the two, harden against rigid is \CsevenSignRigid{} (exact sign test, $p=\CsevenSignPRigid$). The pre-registered tests were one-sided; we report two-sided $p$-values throughout, which are at most twice the one-sided ones. Simulated annealing improved most from tuning for the lowest point; perturbation--compression stays far behind at this budget, partly because its best run was within reach of tightening on only \EligPc{} of instances (against \EligHarden{} to \EligRigid{} for the gradient paths; Table~\ref{tab:diag}). The gate cannot explain rigid's lead over hardening, since both were eligible on at least \EligHarden{} of instances. To check whether it decides the rest of the comparison, campaign C14 tightened the three best runs of every method on the \RefN{} instances fixed in advance for C13 ($n=11,17,23,29$ per family; hexagons $n=8,12,16,20$), with no eligibility condition. Under equal tightening the baselines gain a few instances and the order on these \RefN{} instances, where hardening happens to lead, is unchanged (Table~\ref{tab:refine}); the subset is too small to rule out larger changes for sa and pc elsewhere. without any tightening the order is different, so comparisons of penalty methods on legalised sizes alone would mislead.

\begin{table}[t]\centering\small
\begin{tabular}{lccccc}\toprule
treatment & harden & grow & rigid & sa & pc\\\midrule
no tightening & \RefRawHarden & \RefRawGrow & \RefRawRigid & \RefRawSa & \RefRawPc\\
C07 policy (within 2\%, up to 3 runs) & \RefPolHarden & \RefPolGrow & \RefPolRigid & \RefPolSa & \RefPolPc\\
equal (best 3 runs of every method) & \RefEqHarden & \RefEqGrow & \RefEqRigid & \RefEqSa & \RefEqPc\\\bottomrule
\end{tabular}
\caption{Instances on which each method's lowest size is the lowest of all five, on the \RefN{} instances of C14, under three tightening treatments of the same C07 runs.}\label{tab:refine}
\end{table} No run went below a published value, and \CsevenUnsolved{} instances were solved by no method. The median-tuned comparison (C04, Table~\ref{tab:median}) also puts rigid starts first.

\begin{table}[t]\centering\small
\begin{tabular}{lrccccc}
\toprule
family & $N$ & \textbf{harden} & \textbf{grow} & \textbf{rigid} & \textbf{sa} & \textbf{pc} \\
\midrule
squares in square & 28 & \textbf{27 (5)} & 21 & 23 (1) & 21 & 17 \\
squares in circle & 28 & 16 (6) & \textbf{17 (4)} & 15 (3) & 14 (2) & 8 \\
squares in triangle & 29 & 28 & 27 & \textbf{29} & 24 & 8 \\
triangles in triangle & 28 & 18 & 24 (2) & \textbf{26 (4)} & 19 & 10 \\
triangles in square & 28 & 14 (2) & 17 (3) & \textbf{22 (7)} & 9 & 5 \\
hexagons in square & 18 & 15 & 14 & \textbf{16 (1)} & 14 (2) & 9 \\
circles in square (control) & 29 & \textbf{24} & \textbf{24} & \textbf{24} & \textbf{24 (2)} & 15 (1) \\
\midrule
all & 188 & 142 (13) & 144 (9) & \textbf{155 (16)} & 125 (6) & 72 (1) \\
reached best known & & 125 & 129 & 133 & 115 & 71 \\
\bottomrule
\end{tabular}
\caption{Lowest point by family (C07; every method tuned for its lowest point, 32 runs per instance, matched budget). Each cell: number of instances on which the method\textquotesingle s lowest tightened size is the lowest of all five methods; in brackets, on how many it is the only method that low. Bold: most per family.}
\label{tab:main}
\end{table}

\begin{table}[t]\centering\small
\begin{tabular}{lrccccc}\toprule
$n$ & instances & harden & grow & rigid & sa & pc\\\midrule
2--10 & \RangeSmallN & \RangeSmallHarden & \RangeSmallGrow & \RangeSmallRigid & \RangeSmallSa & \RangeSmallPc\\
11--20 & \RangeMidN & \RangeMidHarden & \RangeMidGrow & \RangeMidRigid & \RangeMidSa & \RangeMidPc\\
21--30 & \RangeLargeN & \RangeLargeHarden & \RangeLargeGrow & \RangeLargeRigid & \RangeLargeSa & \RangeLargePc\\\bottomrule
\end{tabular}
\caption{Instances on which each method's lowest size is the lowest of all five, by $n$ (C07). Small instances are mostly ties; differences open up from $n\approx11$. Rigid starts lead most at $n=21$--30, nearest the $n>30$ instances used for lowest-point tuning, so part of that lead may be a tuning-domain effect; the median-tuned comparison (C04, Table~\ref{tab:median}) also puts rigid starts first.}\label{tab:range}
\end{table}

\begin{figure}[t]\centering
\includegraphics[width=\linewidth]{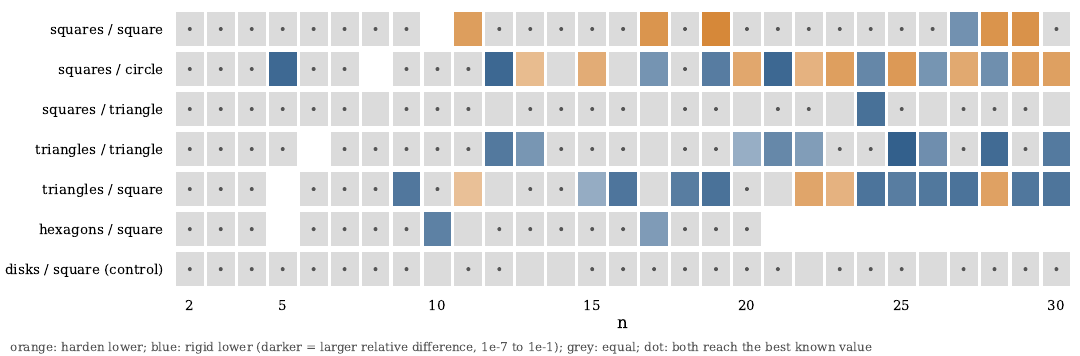}
\caption{Hardening against rigid starts on every instance (C07): which of the two reaches the lower size, and by how much. Blank cells were held out for tuning.}\label{fig:heatmap}
\end{figure}

\begin{figure}[t]\centering
\includegraphics[width=\linewidth]{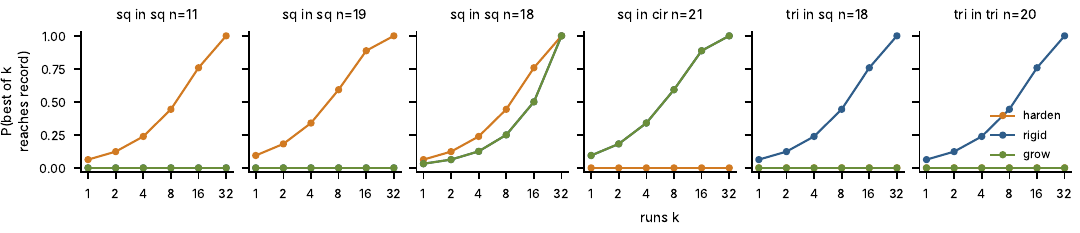}
\caption{Share of $k$-run subsets of the observed 32 runs (C07) that include a run reaching the best known value, $1-\binom{32-s}{k}/\binom{32}{k}$ for $s$ successful runs, on six instances where the gradient paths differ. This is a without-replacement estimate from one pool; it is 1 at $k=32$ whenever $s>0$. Runs not selected for tightening count with their legalised size.}\label{fig:bestofk}
\end{figure}

\paragraph{How often does a method get there?} A count of instances hides how close each call was. Figure~\ref{fig:heatmap} shows every instance: most instances are ties; squares in a square is the only family where every instance but one that differs favours hardening, triangles lean towards rigid starts, and squares in a circle alternate. Figure~\ref{fig:bestofk} shows, for six instances, the share of $k$-run subsets of the observed 32 runs that include a run reaching the best known value. Hardening reaches Trump's $n=11$ and Wainwright's $n=19$ packings in 2 and 3 of 32 runs, and rigid starts in none; on triangle instances the picture is reversed. At $k=32$ the curve is 1 whenever any run succeeded, which says nothing about a fresh batch. Repeated batches do: on the \BatN{} instances of C13, harden and rigid each have three independent 32-run batches at the base budget, and hardening reaches Trump's $n=11$ packing in \BatSqElevenH{} of 3 batches and Schadt's $n=29$ packing in \BatSqTwentyNineH{} of 3, rigid starts in \BatSqElevenR{} and \BatSqTwentyNineR; on 29 triangles in a triangle it is the other way round (hardening \BatTriTwentyNineH{} of 3, rigid \BatTriTwentyNineR). Over all \BatN{} instances, the best known value is reached in \BatHReached{} of the $3\times\BatN$ hardening batches and \BatRReached{} of the rigid ones, and on \BatHSome{} (harden) and \BatRSome{} (rigid) instances some batches reach it and others do not. Statements that one method is ``the only one'' to reach a packing are statements about one batch of 32 runs.

\section{What we learned}\label{sec:learned}
\paragraph{Where hardening differs.} For squares in a square, hardening's lowest size is the lowest of all methods on \CsevenSquInSquHardenLowest{} of \CsevenSquInSquN{} instances and the only one that low on \CsevenSquInSquHardenSole{}; with identical settings, rigid starts manage \CsevenSquInSquRigidLowest{}. In 32 runs each, hardening alone reaches Trump's $n=11$ and Wainwright's $n=19$ packings (Figure~\ref{fig:strip}). For triangles the order reverses: rigid starts are lowest on \CsevenTriInSquRigidLowest{} instances of triangles in a square against \CsevenTriInSquHardenLowest{} for hardening, and \CsevenTriInTriRigidLowest{} against \CsevenTriInTriHardenLowest{} in a triangle (Figure~\ref{fig:tristrip}). On the control family the gradient paths reach the lowest size on the same number of instances, as they should.

\paragraph{With more time on squares.} Campaign C16, a targeted follow-up planned after C07 and C09, tested at ten times the base budget the advantage already seen on squares in a square ($n=17$--$30$, 16 runs per method; Table~\ref{tab:long}). Hardening's lowest size is the lowest of the five on \LongHardenLowest{} of \LongN{} instances (annealing \LongSaLowest, rigid \LongRigidLowest, grow \LongGrowLowest, pc \LongPcLowest), and it reaches the best known value on \LongHardenReached{} (annealing \LongSaReached, rigid \LongRigidReached). It is the only method to reach the best known packing for $n=\LongHardenOnlyReachedNs$; annealing is the only one for $n=\LongSaOnlyReachedNs$. Tightening the best run of every method at the four hardening-only instances, regardless of the tightening gate, leaves every other method short of the best known value (\texttt{scripts/c16\_equal\_tighten.py}). Against rigid starts hardening is lower on \LongHLower{} instances and higher on \LongRLower{} ($p=\LongP$), but the hardening and rigid runs at $n=19$, 28 and 29 repeat C09's exactly (same seeds and budget, deterministic runs), and all three are hardening wins; on the remaining \LongNewN{} instances the count is \LongNewHLower{} against \LongNewRLower{} ($p=\LongNewP$). C16 adds new successes against all four other methods at \LongHardenOnlyReachedNewNs rather than six independent confirmations.

\IfFileExists{tables/C16.tex}{\begin{table}[t]\centering\small
\begin{tabular}{rrccccc}
\toprule
$n$ & best known & harden & grow & rigid & sa & pc \\
\midrule
17 & 4.67553 & \textbf{\checkmark} & 0.324 & 0.032 & 0.032 & 0.092 \\
18 & 4.82288 & \textbf{\checkmark} & 0.177 & \textbf{\checkmark} & \textbf{\checkmark} & 0.015 \\
19$^\dagger$ & 4.88562 & \textbf{\checkmark} & 0.114 & 0.114 & 0.114 & 0.114 \\
20 & 5.00000 & \textbf{\checkmark} & \textbf{\checkmark} & \textbf{\checkmark} & \textbf{\checkmark} & \textbf{\checkmark} \\
21 & 5.00000 & \textbf{\checkmark} & \textbf{\checkmark} & \textbf{\checkmark} & \textbf{\checkmark} & \textbf{\checkmark} \\
22 & 5.00000 & \textbf{\checkmark} & \textbf{\checkmark} & \textbf{\checkmark} & \textbf{\checkmark} & \textbf{\checkmark} \\
23 & 5.00000 & \textbf{\checkmark} & \textbf{\checkmark} & \textbf{\checkmark} & \textbf{\checkmark} & \textbf{\checkmark} \\
24 & 5.00000 & \textbf{\checkmark} & \textbf{\checkmark} & \textbf{\checkmark} & \textbf{\checkmark} & \textbf{\checkmark} \\
25 & 5.00000 & \textbf{\checkmark} & \textbf{\checkmark} & \textbf{\checkmark} & \textbf{\checkmark} & 0.756 \\
26 & 5.62132 & \textbf{\checkmark} & 0.379 & 0.379 & 0.086 & 0.202 \\
27 & 5.70711 & 0.098 & 0.293 & 0.139 & \textbf{\checkmark} & 0.294 \\
28$^\dagger$ & 5.82444 & \textbf{0.042} & 0.176 & 0.073 & 0.178 & 0.179 \\
29$^\dagger$ & 5.93383 & \textbf{\checkmark} & 0.066 & 0.066 & 0.066 & 0.066 \\
30 & 6.00000 & \textbf{\checkmark} & \textbf{\checkmark} & \textbf{\checkmark} & \textbf{\checkmark} & \textbf{\checkmark} \\
\bottomrule
\end{tabular}
\caption{Squares in a square at ten times the base budget (C16, 16 runs per method). \checkmark: reaches the best known value; otherwise the gap to it. Bold: lowest of the five. $^\dagger$: hardening and rigid runs repeat C09 exactly (same seeds and budget, deterministic runs).}
\label{tab:long}
\end{table}
}{}

\paragraph{Separating the disk start from the gradual path.} Hardening differs from rigid starts in two ways: pieces are compressed as disks, and they then harden gradually. Campaign C11 separates the two with \emph{snap}, which compresses disks exactly as hardening does and then switches to the polygons in one step, with the same seeds, settings and budget (Table~\ref{tab:ablate}). Its pre-registered test compares, per family, the instances that exactly one of harden and snap solves. On squares in a square only two instances are discordant, so the test is inconclusive ($p=0.5$). The plan's decision rule then attributed hardening's advantage there to the disk-compressed start. That rule was wrong: failing to detect a difference does not show that there is none, still less which mechanism is responsible. We record the rule and its output, and draw no causal conclusion from it.

Exploratory comparisons of the sizes reached, which were not the pre-registered test, are as follows. We detect no difference between snap and rigid starts (each lower on \SnapRXLower{} and \SnapRRLower{} of \CsevenInstances{} instances). Hardening and snap differ in opposite directions by family: counting instances where hardening is lower against those where snap is lower, squares in a square give \SnapSquSqu{}, triangles in a triangle \SnapTriTri{} and triangles in a square \SnapTriSqu{} (all families, \SnapHLower{} against \SnapXLower, $p=\SnapP$). A gradual path from the same start therefore changes which packings are found; a confirmatory test of the direction on squares in a square would need many more discordant instances than these 28 provide.

Campaign C12 asks the same of area. A hardening piece also grows, by up to 40\% of its area; \emph{grow-area} keeps rigid polygons but scales them so their area follows hardening's schedule exactly. Again we detect no difference from rigid starts (lower on \AreaRXLower{} and \AreaRRLower{} instances), and hardening differs from it in the same directions as from snap: \AreaSquSqu{} on squares in a square and \AreaTriTri{} on triangles in a triangle (all families \AreaHLower{} against \AreaXLower, $p=\AreaP$). Its pre-registered test is again inconclusive on squares in a square. Neither the disk start nor the area schedule alone reproduces hardening's pattern of results.

\begin{table}[t]\centering\small
\begin{tabular}{lrcccc}
\toprule
family & $N$ & harden : snap & snap : rigid & harden : grow-area & grow-area : rigid \\
\midrule
squares in square & 28 & 6 : 0 & 1 : 2 & 5 : 1 & 2 : 1 \\
squares in circle & 28 & 5 : 11 & 10 : 6 & 9 : 8 & 8 : 4 \\
squares in triangle & 29 & 0 : 1 & 0 : 0 & 0 : 1 & 0 : 0 \\
triangles in triangle & 28 & 0 : 8 & 0 : 2 & 1 : 8 & 0 : 3 \\
triangles in square & 28 & 6 : 14 & 8 : 9 & 5 : 11 & 5 : 6 \\
hexagons in square & 18 & 0 : 2 & 1 : 1 & 0 : 2 & 0 : 0 \\
disks in square & 29 & 0 : 0 & 0 : 0 & 0 : 0 & 0 : 0 \\
\midrule all & 188 & 17 : 36 & 20 : 20 & 20 : 31 & 15 : 14 \\
\bottomrule
\end{tabular}
\caption{Which part of hardening matters (C11, C12; harden and rigid from C07, same seeds, settings and budget). Each cell $a:b$ counts the instances on which the first method\textquotesingle s lowest size is lower than the second\textquotesingle s, and the reverse; the rest are ties. \emph{snap} compresses disks as hardening does, then switches to polygons at once; \emph{grow-area} keeps rigid polygons whose area follows hardening\textquotesingle s.}
\label{tab:ablate}
\end{table}

\begin{figure}[t]\centering
\includegraphics[width=\linewidth]{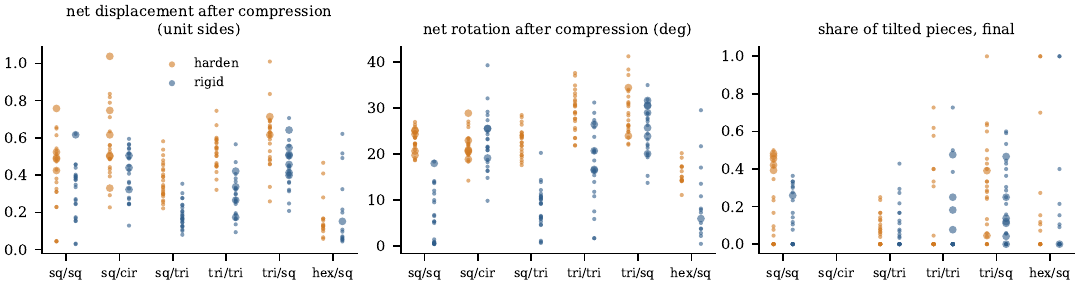}
\caption{What pieces do after compression, on the best run of each C07 instance with $n\ge6$ (orange: hardening; blue: rigid; large dots: the method alone reached the lowest size). Left: mean net displacement; middle: mean net rotation, modulo the piece's symmetry (for hardening this includes turning while pieces are still nearly round and orientation costs almost nothing); right: share of pieces with no edge within $2^\circ$ of a container edge (square and triangle containers).}\label{fig:mechanism}
\end{figure}

\paragraph{What the pieces do.} Figure~\ref{fig:mechanism} measures three things on the replay of each instance's best run (lowest legalised size, before tightening), for hardening and rigid starts: net displacement of pieces between the end of compression and the end of the run, net rotation over the same interval, and the share of pieces that end tilted against every container edge. Midway through the morph phase, where a hardening piece has $\tau\approx0.5$, the orientational order of squares in a square, $|\langle e^{4i\theta}\rangle|$, has a median of \MechSquSquHardenOrder{}; at the same step of a rigid run it is \MechSquSquRigidOrder, so rigid squares are more aligned at that point than hardening squares. On the \MechSoleN{} squares-in-square instances where hardening alone reaches the lowest size, its final packing has more tilted squares than rigid's best on the same instance every time (\MechSoleMore{} of \MechSoleN; median shares \MechSoleTiltH{} against \MechSoleTiltR), as the classical packings at $n=11$, 17 and 19 require. On triangles in a triangle hardening pieces move about twice as far after compression as rigid ones (median \MechTriTriHardenMove{} against \MechTriTriRigidMove{} unit sides), and on the \MechTriFailN{} instances where rigid starts alone reach the lowest size the gap is the same (\MechTriFailMoveH{} against \MechTriFailMoveR). These are descriptive. They fit two hypotheses: that the support function's factor $1-\tau$ lets hardening defer the choice of orientation until neighbours are in place, which squares need when a block must tilt against aligned neighbours; and that triangles, whose corners stick out twice as far as their inradius ($R/\rho=2$, against $\sqrt2$ for a square) and whose good packings alternate up and down, pay for that deferral with large rearrangements. The area explanation alone (a triangle's inscribed disk covers 60\% of it, a square's 79\%) is weakened by hexagons, whose disk covers 91\% and where hardening is no better than rigid starts (\CsevenHexInSquHardenLowest{} against \CsevenHexInSquRigidLowest{} of \CsevenHexInSquN{}, mostly ties). Changing the path also changes contact geometry and forces, so none of this isolates orientation as the cause.

\paragraph{Does choosing the path transfer to new families?} If the better path depends on the family, a search could choose it per instance. Campaign C15 tests that as a decision on four families never used to develop, tune or analyse any method: pentagons and octagons in a square, hexagons in a triangle and triangles in a circle, $n=6$--$20$ (\DecN{} instances). Each instance gets 64 hardening and 64 rigid runs, from which three pre-registered policies of 64 runs are formed: always rigid, always harden, and a pilot that spends 8 runs on each path and the remaining 48 on the path whose pilot reached the lower legalised (untightened) size (Table~\ref{tab:decision}). The pilot and always-rigid rarely differ (lower on \DecPilotLower{} and \DecRigidLower{} instances, $p=\DecP$), so by the plan's rule we report no detected benefit from choosing the path. Always-harden is worse than always-rigid (\DecHLower{} against \DecHRLower, $p=\DecHP$), most clearly on pentagons, and reaches the best known value on \DecReachedHarden{} instances against \DecReachedRigid{} for rigid starts and \DecReachedPilot{} for the pilot. The pilot picked hardening on \DecPicked{} instances and so avoided most of its losses, but was lower than always-rigid on only \DecPilotLower{} instances. The pilot compares untightened sizes, which Table~\ref{tab:refine} shows can misrank methods; in a post-hoc check that compares the tightened best pilot run of each path instead, the pilot picks hardening on \DecTPicked{} instances rather than \DecPicked, changes \DecTChanged{} decisions, and still differs from always-rigid on \DecTLower{} against \DecTRLower{} instances (\texttt{scripts/c15\_tightened\_pilot.py}). None of these families shows a family-level advantage for hardening (it is lower on \DecHLower{} individual instances), so the test says that this pilot policy found no gain on these families; it is not a test of whether a selector can find a favourable family, since none was present.

\IfFileExists{tables/C15.tex}{\begin{table}[t]\centering\small
\begin{tabular}{lrcccc}
\toprule
family & $N$ & harden : rigid & pilot : rigid & pilot chose harden & reached (r / h / p) \\
\midrule
pentagons in square & 15 & 1 : 8 & 0 : 1 & 3 & 8 / 4 / 8 \\
octagons in square & 15 & 0 : 3 & 0 : 0 & 6 & 11 / 9 / 11 \\
hexagons in triangle & 15 & 0 : 2 & 0 : 1 & 8 & 10 / 9 / 10 \\
triangles in circle & 15 & 2 : 4 & 2 : 1 & 3 & 11 / 10 / 12 \\
\midrule
all & 60 & 3 : 17 & 2 : 3 & 20 & 40 / 32 / 41 \\
\bottomrule
\end{tabular}
\caption{Held-out families (C15), $n=6$--$20$, 64 runs per policy. $a:b$ counts instances where the first policy\textquotesingle s result is lower than the second\textquotesingle s and the reverse. The pilot spends 8 runs on each path, then 48 on the one whose pilot went lower. Reached (r / h / p): instances where always-rigid, always-harden and the pilot reach the best known value.}
\label{tab:decision}
\end{table}
}{}

\paragraph{Partly soft starts.} Varying only the starting softness $\tau_0$ on the \CtauN{} most contested instances (Table~\ref{tab:tau}), the counts of instances where each value reaches the lowest size are \CtauOne, \CtauHalf, \CtauQuarter{} and \CtauZero{} for $\tau_0=1, 0.5, 0.25, 0$. Squares in a square favour soft starts (\CtauSquares{}); triangles favour rigid or slightly rounded ones (\CtauTriangles{}). These instances were selected where hardening and rigid starts disagreed, mostly in rigid's favour, which tilts this count toward small $\tau_0$; partly rounded starts are a candidate default, not an established one.

\begin{table}[t]\centering\small
\begin{tabular}{lrrrrr}
\toprule
family & instances & $\tau_0=1$ (harden) & $0.5$ & $0.25$ & $0$ (rigid)\\
\midrule
squares in square & 6 & 3 & 4 & 2 & 0\\
squares in circle & 13 & 2 & 3 & 6 & 5\\
squares in triangle & 1 & 0 & 1 & 1 & 1\\
triangles in triangle & 7 & 0 & 4 & 3 & 5\\
triangles in square & 12 & 2 & 3 & 5 & 5\\
hexagons in square & 1 & 0 & 0 & 1 & 1\\
\midrule
all & 40 & 7 & 15 & 18 & 17\\
\bottomrule
\end{tabular}
\caption{How soft the start must be (C08): on the 40 most contested instances of C07, the number of instances where each starting softness $\tau_0$ reaches the lowest size (32 runs each, all other settings equal; ties counted for every tied value).}
\label{tab:tau}
\end{table}

\paragraph{Longer runs, or more runs?} An earlier test (C09, appendix) suggested that hardening gains more than rigid starts from longer schedules, but it changed run length and restart count together and used instances chosen because the two paths disagreed. Campaign C13 repeats it cleanly on \BudN{} instances fixed in advance ($n=11,17,23,29$ per family; hexagons $n=8,12,16,20$), with 32 runs at three times the budget, and 96 runs at the base budget for the same total cost. Counting instances where hardening is lower against those where rigid starts are lower: \BudOneH{} against \BudOneR{} with 32 base runs ($p=\BudOneP$), \BudThreeH{} against \BudThreeR{} with 32 long runs, and \BudMoreH{} against \BudMoreR{} with 96 base runs. Longer runs improved rigid starts more than hardening (rigid lower than its base on \BudGainR{} instances and higher on \BudLossR; hardening \BudGainH{} and \BudLossH), and at equal cost, three times longer runs and three times as many runs did about equally well for both paths (\BudLvMLH{} against \BudLvMMH{} for hardening, \BudLvMLR{} against \BudLvMMR{} for rigid). The C09 result does not replicate; we withdraw it.

\paragraph{Two paths, or more runs of one?} The paths fail on different families, so running both might help. At equal total budget, 32 rigid runs plus 32 hardening runs solve \PortRigidHarden{} of the \CsevenInstances{} instances, against \PortRigidSixtyFour{} for 64 rigid runs and \PortRigidGrow{} for 32 rigid plus 32 growing runs (instances solved by exactly one of the first two: \PortSign, two-sided $p=\PortP$). On squares in a square alone the mixed portfolio solves one instance more than doubled rigid runs: it gains $n=11$ and 19 and loses $n=27$. We detected no benefit from mixing; at this size that does not show there is none.

\paragraph{Tuning decides comparisons.} Our first experiments compared hardening with rigid starts that inherited the hardening schedule, on two square instances, and hardening looked better. Tuned with comparable effort, rigid starts led overall. Re-tuning for the lowest point instead of the median changed four of the five methods' settings (stronger pressure for the gradient paths, more noise for rigid starts) and helped simulated annealing most. It kept the broad pattern, hardening for squares in a square and rigid or growing starts for triangles, though the leading method changed for squares in a circle and triangles in a square.

\section{When to use it, and downstream use}\label{sec:matters}
\paragraph{When to add hardening.} Rigid starts remain the better default. Hardening is worth keeping beside them on squares in a square, where it reached packings none of the other four methods reached (Section~\ref{sec:learned}) and where the methods cover different instances: at ten times the budget, hardening and annealing together reach every best known value from $n=17$ to 30 except $n=28$. Our hypothesis is that it helps where good packings tilt a block of pieces against an otherwise aligned arrangement, but that is not a validated predictor: we have no test that identifies favourable families in advance, and we have not shown that splitting a fixed budget between hardening and rigid starts beats spending it on rigid starts (C10). On triangles and on the held-out families it should not replace rigid starts.

\paragraph{Downstream use.} The earlier cube release \cite{softtorigidpacking} has been reused by others; we report this separately from our results, which it does not change. A continuation of the unfinished $n=13$ cube search \cite{n13repo} builds on our simulator, reuses our tightening code and checkers unchanged, adds two moves (re-melting a finished packing: soften, expand slightly, reharden; and inserting a thirteenth cube into the 12-cube packing), and after 2{,}627 runs reports a certified packing of side $2.956149$ at $10^{-6}$ clearance, consistent with Friedman's $2.956+$ and not claimed as an improvement; C05 reaches a comparable numerical contact value ($2.956145$) without clearance. Zarzuelo \cite{platonicrepo} generalises the construction to all five Platonic solids as pieces and containers, with exact feasibility checks over $\mathbb{Q}(\sqrt2,\sqrt5)$ and an archive search that repeatedly softens and rehardens existing packings; the reported values are certified upper bounds, and comparisons with Berthold et al.\ \cite{Berthold2026} must convert between edge and circumradius normalisations. Yudhiono \cite{tesseractrepo} reimplements ball-to-hypercube continuation independently in four dimensions and combines it with structured seeds, constrained polishing and basin hopping, reporting floating-point candidates for 26--30 tesseracts. Together they continue a partially explored search, extend the geometry and the exact verification to new solids, and reimplement the method in a higher dimension. None is a controlled comparison, and the first two inherit our simulator (one modified, one generalised) and tightening scheme, so they are not independent validations of it; the generalised exact checker is new verification work. What they add is a use we did not test: softening as a rearrangement operator inside a larger search rather than only as a cold start. Whether temporary softening escapes a rigid local optimum better than rigid perturbation at equal budget, from shared starting packings, is the experiment they motivate.

\paragraph{Other settings.} The same question arises wherever an overlap penalty is minimised in a shrinking container: irregular nesting \cite{Wascher2007,Bennell2008,Leao2020}, where inscribed circles are already an overlap proxy \cite{Jones2014,Gardeyn2025}; floorplanning, where circles-then-rectangles is an established two-stage switch \cite{AnjosVannelli2002,Luo2008}; and faceted particles in confinement \cite{Torquato2018,Wang2018}. In each, C11 is the test to run first, and real objectives add stability, orientation and ordering constraints \cite{Bortfeldt2013,Araujo2019} that any gain in density must survive.

\section{Limitations and open directions}\label{sec:limits}
\paragraph{Limitations.} These are search results, not optimality proofs, and tightening is a local method. In the main comparison (C07) each instance had 32 runs per method at a fixed budget, and C16 had 16; the $n=12$ cube record came from one of 192 runs, and ``only one method reached it'' holds within the run count and budget stated for that campaign. Gradient paths are matched in steps, not pair evaluations or wall-clock time (hardening was the fastest per run on polygon families); the baselines are matched to a hardening run's pair evaluations, which perturbation--compression overshoots (Table~\ref{tab:diag}). Simulated annealing and perturbation--compression are our implementations; their best runs were less often close enough to be tightened, so the tightening policy favours the gradient paths, most for pc (C14 shows the order survives equal tightening on 28 instances); in C16 the other methods' best runs at the hardening-only instances were tightened after the fact and remained short of the best known value. The lowest-point tuning used instances larger than the test set, and in that round every winner sat on a grid edge or corner and, as pre-registered, the grid was not extended; simulated annealing moved furthest between the two tuning rounds, so its results are best read as a lower bound on what a wider grid would give. Its step size adapts every 200 moves to keep the acceptance rate between 20\% and 40\%, so at $T_0=10^{-3}$ it behaves as a Metropolis walk with small adaptive steps rather than one with large uphill moves. The grids' ranges were first chosen during development on squares in a square, and settings suited to triangles could lie outside them; the C08 sweep of starting softness is only a partial check. In 3D only the three gradient paths were compared. Catalogue values come from captions and witnesses as of October 2026; truncated values are only known to their printed precision. Not detecting a difference (between mixed and single-path portfolios, between snap or grow-area and rigid starts, or between the pilot and always-rigid) is not evidence of equivalence at these sample sizes.

\paragraph{Open directions.}
\begin{itemize}\itemsep1pt
\item \emph{A rule that predicts the path.} The pilot showed no detected benefit on held-out families that had no aggregate hardening advantage. A useful rule would have to predict, before running, the families where it does; we have one such family and no rule that singles it out in advance.
\item \emph{External baselines.} Our simulated annealing and perturbation--compression are our own implementations. Running a published nester \cite{Gardeyn2025} or solver \cite{Berthold2026} once, untouched, at the same budget would test whether they were given a fair fight.
\item \emph{Softening as a move.} Re-softening an existing packing and rehardening it, as downstream searches do \cite{n13repo,platonicrepo}, against rigid perturbation from the same packings at equal budget.
\item \emph{Other paths.} Paths through other shape families (an ellipse for a rectangle, a rounded octahedron for a cube, superellipses \cite{Jiao2008}), paths that harden pieces one at a time, or schedules that adapt $\tau$ to the current overlap.
\item \emph{Larger $n$, other solids, other solvers.} Most unsolved instances have $n\ge17$. Our controlled study did not include solids other than cubes; downstream work now explores them \cite{platonicrepo} and higher dimensions \cite{tesseractrepo}, without controlled baselines. Hardening could also seed exact models \cite{Berthold2026} or shrink-and-separate nesters \cite{Gardeyn2025}.
\item \emph{Non-convex and mixed pieces.} The rounded-shape construction needs a convex core; non-convex pieces would need a different family of intermediate shapes, for example a convex decomposition hardened part by part.
\end{itemize}

\sloppy
\paragraph{Reproducibility.} Every run, replay, certificate and decision is in the repository \url{https://github.com/yoheinakajima/soft-to-rigid} and its site, \url{https://yoheinakajima.github.io/soft-to-rigid/}. Each experiment was planned in writing before it ran; the plans, the raw results and the research ledger (an ActiveGraph event log \cite{activegraph}) are committed, and every number in this paper is generated from the ledger.

\paragraph{Acknowledgements.} Code, experiments and verification were developed with Claude (Anthropic), an AI system, under the author's direction. The author takes responsibility for the content.

\FloatBarrier
\bibliographystyle{abbrvnat}
\setlength{\bibsep}{1pt plus 0.3ex}
\renewcommand{\bibfont}{\footnotesize}
\bibliography{refs}

\clearpage
\appendix
\section{Supplementary material}
\suppressfloats[t]

\paragraph{Campaigns.} C00 reproduce the earlier square results; C01--C03 median tuning; C04 breadth (median-tuned); C05 cubes; C06 lowest-point tuning; C07 breadth (lowest-point-tuned, main); C08 starting softness; C09 longer schedules (confounded); C10 portfolio; C11 snap; C12 grow-area; C13 clean budget study; C14 equal tightening; C15 held-out families (path choice); C16 squares in a square at ten times the budget. Plans, amendments, raw runs, reports and the decision recorded for each are in \texttt{campaigns/} and the journal.

\paragraph{Frozen settings.} Median-tuned (C01--C03): harden $\mu=0.08$, noise $\times8$; grow $\mu=0.04$, $\gamma_0=0.2$; rigid $\mu=0.04$, noise $\times2$; sa $T_0=3\cdot10^{-3}$, $\mu=0.0025$; pc kick $0.05$, relax $19200$ steps. Lowest-point-tuned (C06): harden and rigid $\mu=0.16$, noise $\times8$; grow $\mu=0.08$, $\gamma_0=0.1$; sa $T_0=10^{-3}$, $\mu=0.02$; pc kick $0.05$, relax $19200$ steps. All are in \texttt{methods/*.json}, each with the campaign and grid point that selected it.

\paragraph{Longer schedules, first attempt (C09).} C09 gave hardening and rigid starts three and ten times the budget on the \CbudgetN{} instances where the two differed most. It changed run length and restart count together (32 runs at budget 1, 16 at budgets 3 and 10) and selected instances for disagreement, mostly in rigid's favour, so part of its change is regression toward an even split. We report it (Table~\ref{tab:budget}) but draw conclusions about run length only from C13, which fixes these confounds.
\begin{table}[t]\centering\small
\begin{tabular}{rrrrr}
\toprule
budget & harden lowest & rigid lowest & harden reaches best known & rigid reaches best known\\
\midrule
1$\times$ & 3 & 17 & 1 & 10\\
3$\times$ & 5 & 16 & 2 & 10\\
10$\times$ & 9 & 16 & 6 & 9\\
\bottomrule
\end{tabular}
\caption{Longer schedules (C09) on the 20 most contested instances: number of instances where each path reaches the lower of the two lowest sizes (ties counted for both), and where it reaches the best known value. Budget 1 is C07 (32 runs); budgets 3 and 10 use 16 runs.}
\label{tab:budget}
\end{table}

\begin{table}[t]\centering\small
\begin{tabular}{lrccccc}
\toprule
family & $N$ & \textbf{harden} & \textbf{grow} & \textbf{rigid} & \textbf{sa} & \textbf{pc} \\
\midrule
squares in square & 28 & \textbf{28 (7)} & 21 & 20 & 10 & 17 \\
squares in circle & 28 & 16 (7) & 11 (2) & \textbf{19 (9)} & 10 & 8 \\
squares in triangle & 29 & 27 & 24 & \textbf{29 (1)} & 13 & 7 \\
triangles in triangle & 28 & 19 (1) & 22 (3) & \textbf{24 (4)} & 8 & 10 \\
triangles in square & 28 & 10 (1) & \textbf{19 (5)} & 18 (7) & 12 (1) & 5 \\
hexagons in square & 18 & 15 & 15 (1) & \textbf{17 (2)} & 8 & 6 \\
circles in square (control) & 29 & 27 & 27 & \textbf{28} & 11 (1) & 11 \\
\midrule
all & 188 & 142 (16) & 139 (11) & \textbf{155 (23)} & 72 (2) & 64 (0) \\
reached best known & & 121 & 119 & 132 & 69 & 63 \\
\bottomrule
\end{tabular}
\caption{Settings tuned by median gap (C04). Same layout as Table~\ref{tab:main}.}
\label{tab:median}
\end{table}

\begin{table}[ht]\centering\small
\begin{tabular}{lrrrrrrr}
\toprule
method & \multicolumn{2}{c}{steps (sa: moves) per run} & \multicolumn{2}{c}{pair evaluations per run} & \multicolumn{2}{c}{ms per run} & best run \\
 & $n=20$ & $n=30$ & $n=20$ & $n=30$ & $n=20$ & $n=30$ & tightened \\
\midrule
harden & 11{,}500 & 11{,}500 & 418{,}365 & 650{,}532 & 451 & 699 & 97\% \\
grow & 11{,}500 & 11{,}500 & 426{,}807 & 668{,}192 & 727 & 1{,}183 & 98\% \\
rigid & 11{,}500 & 11{,}500 & 348{,}285 & 550{,}494 & 633 & 1{,}001 & 100\% \\
sa & 66{,}854 & 104{,}043 & 448{,}878 & 714{,}120 & 582 & 983 & 94\% \\
pc & 19{,}148 & 20{,}530 & 690{,}285 & 1{,}176{,}434 & 1{,}117 & 1{,}993 & 56\% \\
\bottomrule
\end{tabular}
\caption{Work per run (C07, squares in a square, mean over 32 runs) wall-clock milliseconds per run on one core (Node.js), and the share of instances on which each method's best run was within 2\% of the best known value and so was tightened (all families). Gradient paths are matched in steps; the baselines stop once they exceed a reference hardening run's pair evaluations.}\label{tab:diag}
\end{table}

\begin{figure}[t]\centering
\includegraphics[width=\linewidth]{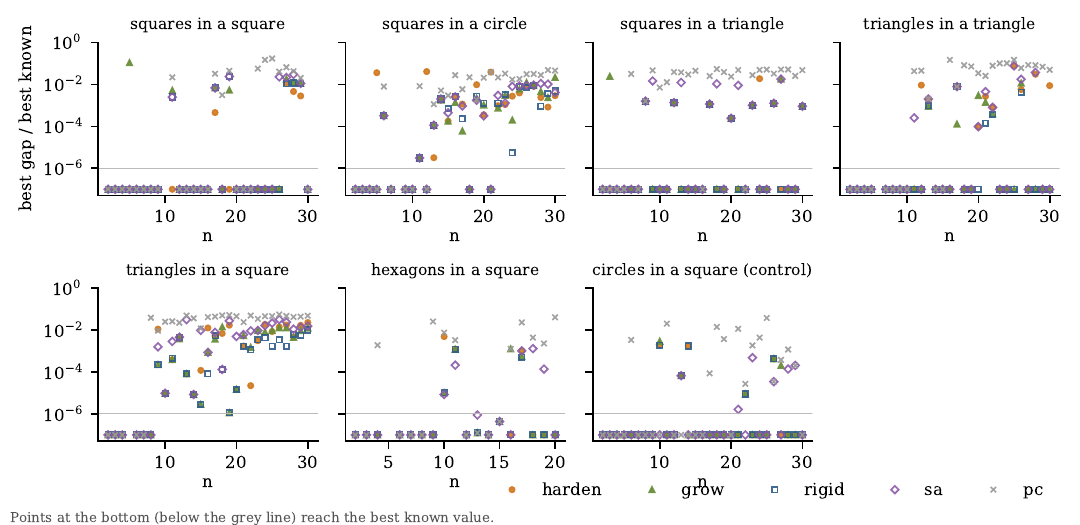}
\caption{Lowest size reached per instance and method, relative to the best known value (C07). Points on the bottom row reach it.}\label{fig:gaps}
\end{figure}

\end{document}